# Characterisation of the signal to noise ratio of 2-photon microscopes

*Radek Macháň[1], Shau Poh Chong[1], Khee Leong Lee[2] & Peter Török[1,3,4*]*

*[1]Singapore Centre for Environmental Life Sciences Engineering (SCELSE), Nanyang Technological University, 60 Nanyang Drive, Singapore 637551, Singapore*
*[2]Institute for Digital Molecular Analytics and Science (IDMxS), 59 Nanyang Drive, Singapore 639798, Singapore*
*[3]Lee Kong Chian School of Medicine, Nanyang Technological University, 59 Nanyang Drive, Singapore 639798, Singapore*
*[4]School of Physical and Mathematical Sciences, Nanyang Technological University, 50 Nanyang Drive, Singapore 639798, Singapore*

**Correspondence: peter.torok@ntu.edu.sg*

**Abstract**

The signal to noise ratio (SNR) is a key performance metric of an imaging instrument. Here we describe the characterisation of SNR of a custom 2-photon microscope and compare the SNR performance of our microscope with selected commercial 2-photon microscopes. The methodology described in this paper can serve as guidance for others wishing to characterise and benchmark their 2-photon or other point-scanning microscopes.

**Introduction**

Signal to noise ratio (SNR) is a critical quality metric of microscope images with impact on the visibility of structures of interest and image resolution [1–3]. The ability to capture images with high SNR is, therefore, a key target of microscope design. In optical microscopy, the maximum achievable SNR is limited by the photon counting noise, also known as the shot noise. This is a fundamental physical limit which originates from the stochastic nature of photon emission and detection events. Consequently, if a stationary photon flux $\varphi$ reaches a photodetector, the number of photoelectrons generated in the active medium of the detector at any time interval $\Delta t$ must follow Poisson statistics with the mean $n_p = \varphi\, \Delta t\, Q$, where $Q$ is the quantum efficiency of the detector, and standard deviation equal to the square root of the mean, $\sigma(n_p) = \sqrt{n_p}$.

Assuming that the response of the detector and all signal processing electronics is linear, the mean signal, i.e. the mean value in analogue-to-digital units (ADU) assigned to the corresponding pixel, is $S = K\, n_p$, where $K$ is a constant describing the detector gain, electronic amplification of the detector output as well as the range and the bit-depth of the signal digitiser, during $\Delta t$ signal integration time (exposure time in the case of widefield detection by a camera or pixel-dwell time in the case of point-scanning microscopy modalities). Note that in some cases the image pixel value $\bar{S}$ may contain a constant offset $B$, in which case it is possible to write $S = \bar{S} - B$ [4]. In the following discussion we will assume that the offset $B$ has been subtracted from the images.

The standard deviation of the signal $S$ is given by

$$\sigma(S) = K\sqrt{n_p + \sigma_R^2}\,, \tag{1}$$

where $\sigma_R^2$ is the readout noise - a photon-flux-independent noise contributed by the detector expressed in the units of electrons. If we define SNR as the ratio of the mean signal of an image pixel and the standard deviation, we can write:

$$SNR = \frac{S}{\sigma(S)} = \frac{\sqrt{n_p}}{\sqrt{1 + \frac{\sigma_R^2}{n_p}}}. \tag{2}$$

The detectors used in this study, and generally common in 2-photon microscopy, were photomultiplier tubes (PMTs) [5]. PMTs are well suited for low light level applications because of their negligible readout noise, and the capability of multiplying the number of photoelectrons emitted from the photocathode. The latter is achieved by accelerating the photoelectrons by high voltage applied over a series of dynodes where each accelerated electron on collision with a dynode produces a shower of secondary electrons. Weak signals can be thus increased to levels significantly above the noise floor. However, the secondary electron emission, being a stochastic process, also contributes to the noise. Unlike in the case of the readout noise, the contribution of the electron multiplication noise cannot be easily separated from the contribution of the shot noise, so it is commonly treated as a reduced apparent quantum efficiency $Q_{app}$ in a perfect shot noise limited detector [2,5]. We can then write for the SNR:

$$SNR = \sqrt{n_{p,app}} = \sqrt{\varphi\, \Delta t\, Q_{app}}\,. \tag{3}$$

It can be seen from equation (3) that the SNR increases by a factor of $\sqrt{k}$ when the integration time is increased $k$ times or when $k$ acquisitions are averaged (see Fig. S1).

The output current from a PMT $I_{in}$ is converted to voltage $V_{out}$ by a transimpedance amplifier (TIA). The voltage signal is then digitised and stored in the computer which controls the microscope. The two main parameters to consider when choosing a TIA are its transimpedance gain ($V_{out}/I_{in}$) and frequency bandwidth [6]. The gain determines how the current input $I_{in}$ is mapped to the voltage output $V_{out}$ of the TIA, which is then the input signal of the digitiser. Too high gain leads to a loss of information due to clipping high signal values, which are translated to voltage values outside the digitiser input range or the TIA output range (saturation). To avoid this, the gain of the PMT would have to be reduced by operating the PMT at a lower accelerating voltage. This may in turn reduce the SNR, as at lower accelerating voltages the signal may not be raised sufficiently above the noise floor. While increasing the gain of the TIA cannot improve the SNR as the gain of the PMT does, it could potentially decrease the SNR through the $Q_{app}$.

The frequency bandwidth of the TIA is especially important when fast beam scanning, e.g. a resonant scanner, is employed. If the maximum frequency supported by the TIA is smaller than $1\,/\Delta t$ with $\Delta t$ being the pixel dwell time, the TIA, acting as a low-pass filter, will effectively perform sliding window averaging of neighbouring pixels along the fast scanning axis. It follows from what we deduced above from equation (3), that the SNR will be thus increased by $\sqrt{k}$, where $k$ is in this case the number of averaged neighbouring pixels. At the same time spatial resolution of the image will be reduced in an anisotropic manner (unless the nominal pixel size is sufficiently small for resolution to be unaffected by its $k$-times increase). The upper limit of the TIA bandwidth is commonly defined as the frequency at which the gain drops to 70.7% of its maximum value (decrease of 3 dB). Depending on the dependence of the TIA gain on the signal frequency, signals at frequencies significantly exceeding the nominal bandwidth may still be transmitted, albeit with reduced gain.

The noise performance of a microscope can be characterised using a temporal series of images of a stationary object. For each individual pixel we obtain a series of values, the average of which can be considered the signal $S$ and the standard deviation of which is a measure of the noise [4,7,8]. According to equation (1), the plot of the variance versus the mean signal $S$ for all pixels is linear (in the absence of saturation) with a slope of $K$. Similarly, we can construct a plot of SNR versus the signal $S$ for each pixel according to equation (2). Note that this approach finds SNR in a pixel-wise manner and does not generate a single number for the whole image (except for the trivial case when all pixels have the same mean value). The points, corresponding to individual pixels, in both types of plots (variance vs. $S$ and SNR vs. $S$) are distributed along curves, which are independent of the imaged object or excitation intensity, but depend on $K$ (i.e. on the PMT and TIA gain as well as the digitiser range, which all contribute to $K$) as shown in Fig. S2. If, instead of versus $S$, we plot these quantities against $S/K = n_{p,app}$, the points follow curves that are entirely independent of the experimental settings (as seen from equation (3) and Fig. S3).

Saturation, caused either by the saturating the detector or, more frequently, by the TIA output voltage exceeding the range of the digitiser, results in an apparent decrease in variance, and consequently an increase in the SNR for the saturated pixels. At very high saturation levels, when for a given pixel all the values in the time series are saturated, variance equals 0 and the SNR defined by equation (2) tends to infinity. While the apparently high SNR values found in saturated pixels may be appealing, they do not improve image quality as demonstrated in Fig. S4. Thus, saturation should not be sought as a means to improving SNR but should be avoided.

Sometimes it is useful to assign a single SNR value to the entire image. The following definition, different from equation (2), is commonly used for the SNR of a whole image:

$$SNR_{Im} = \frac{\sigma_S}{\sigma_n}, \qquad (4)$$

where $\sigma_S$ and $\sigma_n$ are the standard deviation of the signal and the noise, respectively, calculated in this case across all pixels of the image. $SNR_{Im}$ depends not only on the number of photons contributing to the image and the detector performance, but also on the captured scene, as can be readily seen by considering a perfectly uniform image, which would have zero $SNR_{Im}$ according to equation (4) regardless of the number of photons contributing to each pixel. $SNR_{Im}$ can be seen as a proxy for the information content of the image.

It can be also deduced from equation (4) that $SNR_{Im}$ cannot be unambiguously determined from a single image unless some additional information is available on and/or assumptions made about the image [8,9]. If we have a temporal series of images of a given scene consisting of at least two images, we can make use of the assumption that the signal does not change throughout this series and that the noise present in different images is not correlated [9,10].

$SNR_{Im}$ can be then calculated from the cross-correlation function of two images from the series as

$$SNR_{Im} = \sqrt{\frac{G_{ij}}{G_{ij} - 1}} \tag{5}$$

where

$$G_{ij} = \frac{\langle \bar{f}_{ix}\, \bar{f}_{jx} \rangle}{\sqrt{\langle \bar{f}_{ix}^{\,2} \rangle \langle \bar{f}_{jx}^{\,2} \rangle}} \tag{6}$$

is the cross-correlation function of $i$-th and $j$-th image in the series, $f_{ix} = \bar{f}_{ix} + \langle f_{ix} \rangle = s_{ix} + n_{ix}$ is the value of the $x$-th pixel in $i$-th image, $s_{ix}$ and $n_{ix}$ the respective signal and noise and $\langle\, .\, \rangle$ indicate averaging over all pixels.

In this work we have used both the pixel-wise (equation (2)) and image-based (equation (5)) definitions of the SNR to characterise the performance of our custom-built 2-photon microscope (referred to as NOBIC 2PM). In what follows we also compare the performance of the NOBIC 2PM to two commercial systems. In all experiments we imaged samples of fern asparagus (*Asparagus setaceus*) which we found to emit autofluorescence across the entire visible spectrum without any signs of degradation for several weeks and minimal photobleaching even at high excitation intensities (see Fig. S5). We found that both the NOBIC 2PM and our custom designed TIA compared favourably with their commercial counterparts.

## Results and discussion

*Comparison of TIAs*

We have compared 4 different TIAs in the NOBIC 2PM, two versions of our custom TIA (NOBIC TIA) and two commercial ones, one of which had adjustable gain. The gain and the frequency bandwidth of the TIAs we tested are summarised in table 1.

| TIA | Gain / kV/A | Frequency bandwidth / MHz |
|---|---|---|
| NOBIC – high gain | 194.4 | 8.69 |
| NOBIC – low gain | 21.68 | 11.45 |
| Hamamatsu C12419 | 1000 | 1 |
| Femto DHPCA-100 | 10 | 14 |
| | 100 | 3.5 |

**Table 1**: The gain and the frequency bandwidth of the TIAs used in this study. The values for the commercial TIAs are the nominal values provided by the manufacturers; the values for the NOBIC TIA have been measured as described in the Materials and Methods section.

No difference in the SNR was observed between the two versions of the NOBIC TIA (see table 2). Any differences in the observed $SNR_{Im}$ values are more likely attributable to bleaching of the sample or to focus drift during the time needed for exchanging the TIA than to any differences in the respective TIA performance.

| TIA | $SNR_{Im}$ – scene 1 | $SNR_{Im}$ – scene 2 |
|---|---|---|
| NOBIC – high gain | 1.58 ± 0.03 | 1.44 ± 0.03 |
| NOBIC – low gain | 1.47 ± 0.04 | 1.38 ± 0.05 |

**Table 2**: $SNR_{Im}$ obtained with the two versions of the NOBIC TIA for images of two scenes. The values are presented as the mean and the standard deviation from 5 repeated acquisitions. For scene 1, the high-gain version was used before switching to the low-gain one; for scene 2 it was vice-versa. The same PMT voltage and laser power was used throughout all the measurements.

Hamamatsu C12419 achieved significantly higher values of SNR compared to the high-gain version of the NOBIC TIA (see Figs. 1 and S6). Closer inspection of the images acquired with Hamamatsu C12419 reveals that the resolution appears anisotropic with pixels "smeared" along the fast scanning axis (X). This is not surprising considering that the frequency bandwidth of this TIA is insufficient for the short pixel integration time. To quantify the extent of the pixel averaging by the TIA, we calculated the spatial autocorrelation of the images for shifts along the fast (X) and slow (Y) scanning axis. The autocorrelation function for shifts along X axis is given by:

$$G(\Delta X) = \frac{\langle \bar{f}(x,y)\bar{f}(x+\Delta X,y)\rangle}{\sqrt{\langle \bar{f}(x,y)^2\rangle\langle \bar{f}(x+\Delta X,y)^2\rangle}}, \tag{7}$$

and, analogously, $G(\Delta Y)$ was calculated for shifts along Y axis. $\bar{f}(x,y) = f(x,y) - \langle f(x,y)\rangle$ is the difference of a pixel value from the mean pixel value. Figs. 1 B and S6 B compare results for images acquired with Hamamatsu C12419 and with the NOBIC TIA. If the TIA bandwidth is sufficient and each pixel value is independently sampled, the autocorrelation drops abruptly at the shift of a single pixel. This drop is caused by the contribution of the shot noise which is uncorrelated between different pixels. In the case of an insufficient TIA bandwidth, the shot noise contribution to the correlation is spread over several neighbouring pixels and decays at pixel shifts corresponding to the width of the effective averaging sliding window. The remaining correlation is caused by the actual structures in the image and decays gradually at pixel shifts corresponding to the typical dimensions of the imaged structures.

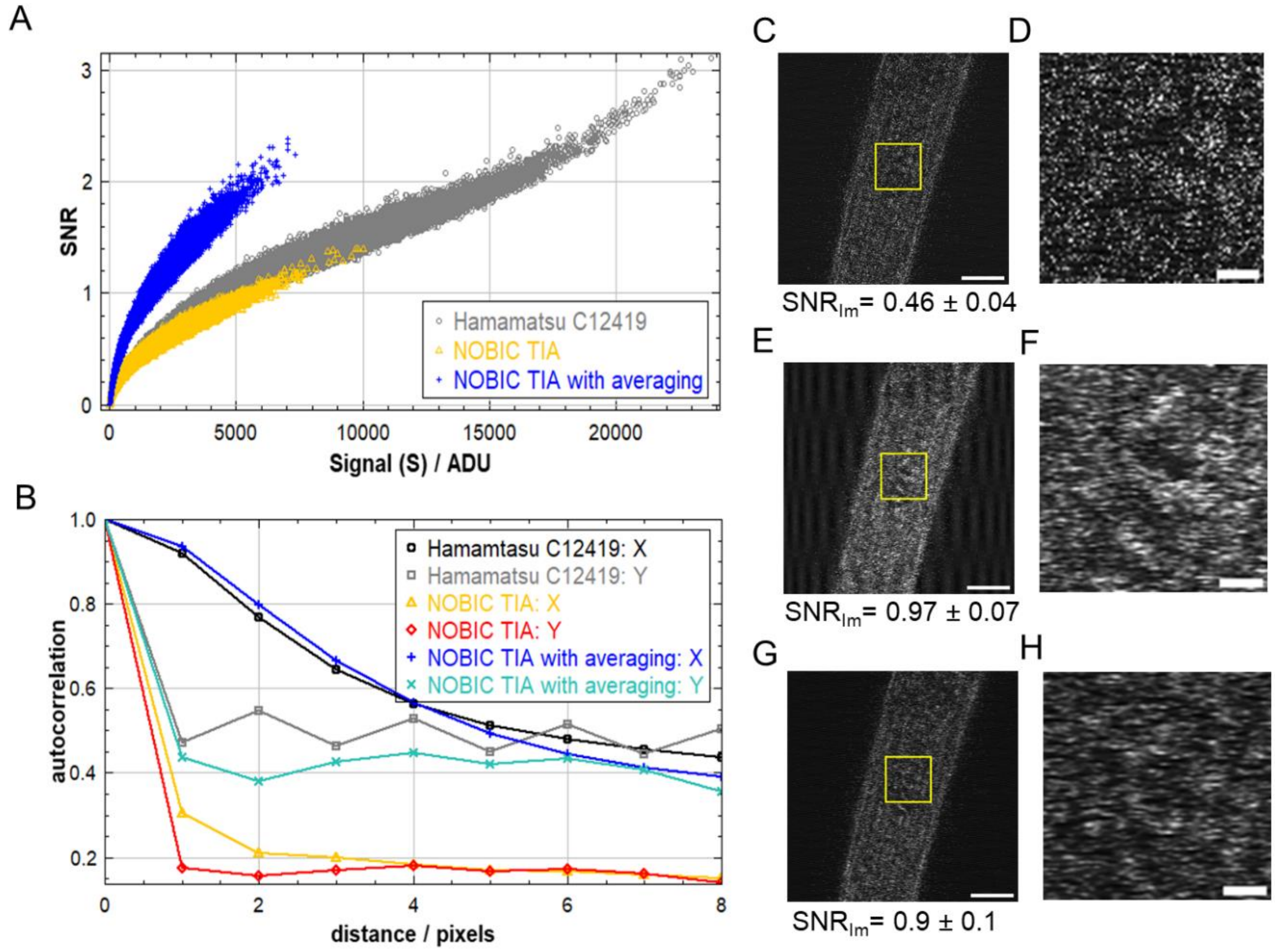

**Fig. 1:** Comparison of the NOBIC TIA (high-gain version) and Hamamatsu C124190: SNR vs. signal plot (A), autocorrelation functions calculated from a single frame in the stack (B) and examples of single frames acquired using the NOBIC TIA (C, D) and Hamamatsu C124190 (E, F), with the corresponding $SNR_{Im}$ (mean ± standard deviation from all pairs of frames in the stack). The scale bar in panels C and E represents 50 µm and the yellow square indicates the area shown magnified in panels D and F, in which the scale bar represents 10 µm. An example of a frame acquired with the NOBIC TIA after averaging along X axis (as described in the text) is shown together with a magnified view (G, H) and data for an image averaged in that way are plotted in panels A and B. The figure reveals that the Hamamatsu TIA does not possess sufficient bandwidth for this application and hence its apparent SNR advantage over the NOBIC TIA is superficial.

If we now consider the case where shot noise contributes to the correlation only at zero pixel shift (which is for example the case of $G(\Delta Y)$ for all images), we notice that the ratio of $G(1)$ to $(G(0) - G(1))$ is proportional to the ratio of the contributions of the signal and the noise to the autocorrelation, therefore the SNR. We explore this further in Fig S7 and show that $SNR_{Corr}$ calculated as

$$SNR_{Corr} = \sqrt{\frac{G(1)}{1 - G(1)}} \tag{8}$$

gives results consistent with $SNR_{Im}$. Note that this does not contradict what was said earlier about the impossibility to unambiguously determine SNR from a single image since it depends on the assumption that the signal (that means the structures) in the image is correlated over lengths exceeding a single pixel and that the signal correlation does not decrease significantly at a shift of a single pixel (or at least that it decreases to a comparable extent in all compared images).

To see, how much the difference in SNR between Hamamatsu C12419 and the NOBIC TIA can be attributed to the difference in their frequency bandwidth, we have modelled the operation of the TIA, by applying averaging along X axis on the images acquired with NOBIC TIA. Figs. 1 B and S6 B show $G(\Delta X)$ results for the NOBIC TIA after convolving the image with a 1-dimensional Gaussian kernel, the width of which was selected to result in $G(\Delta X)$ decaying visually similarly to that of the images taken with Hamamatsu C12419. Figs. 1 A and S6 A show the corresponding SNR results. The SNR results became comparable after the sliding window averaging, indicating that the narrower frequency bandwidth of the Hamamatsu C12419 is sufficient to explain the differences in the SNR.

To estimate the effect of insufficient TIA bandwidth on the SNR, we can approximate the averaging by convolution with a 1-dimensional Gaussian function. This will result in the autocorrelation function in the *X*-axis direction having the form $G(\Delta X) = \bar{G}(\Delta X) + A_N\, e^{-\left(\frac{\Delta X}{d}\right)^2}$, where $\bar{G}(\Delta X)$ is the autocorrelation function of the signal (without any contribution from the noise) and $A_N$ is a scaling factor reflecting the relative contribution of the noise. The distance $d$ is a measure of the extent of the averaging and can be approximately estimated by fitting of the initial part of $G(\Delta X)$, where we can assume $\bar{G}(\Delta X)$ to decay only moderately. The factor by which SNR increases because of the averaging can be then estimated as $\sqrt{d\sqrt{\pi}}$. In the case of the Hamamatsu C12419 shown in Fig. 1 B, the Gaussian fit of the first 4 points of $G(\Delta X)$ yields $d = 1.75$, which translates to approximately twofold increase in the SNR compared to the case without any averaging. This is consistent with the results shown in Fig. 1.

An analogous effect can be also observed when the pixel integration time $\Delta t$ is changed. When using a resonant scanner (as in the case of NOBIC 2PM) the scanning speed is fixed, but changing the number of pixels per line effectively changes the pixel dwell time $\Delta t$.We explore this in Fig. S8. From equation (3) we would expect the SNR to decrease as the square root of $\Delta t$; however, for very short $\Delta t$, i.e. for high pixel frequencies, the effective pixel averaging caused by the limited TIA frequency bandwidth reduces the impact of $\Delta t$ on the SNR.

Next, we compared the high-gain version of the NOBIC TIA with Femto DHPCA-100 operated at two different gains (10 and 100 kV/A). As seen from table 1, the different gain settings of Femto DHPCA-100 also affect its frequency bandwidth. Figs 2 A, B show the SNR results. Femto DHPCA-100 at 100 kV/A gain has the highest SNR, which is probably again a result of an insufficient frequency bandwidth (see Figs 2 C and S9). On the other hand, Femto DHPCA-100 at 10 kV/A gain has the highest frequency bandwidth and slightly lower SNR than the NOBIC TIA. As can be seen from Figs 2 C and S9, $G(\Delta X)$ shows signs of slight pixel averaging even for the NOBIC TIA, however Femto DHPCA-100 at 10 kV/A gain does not show any averaging. Therefore, the difference in SNR is probably again related to the TIA frequency bandwidth.

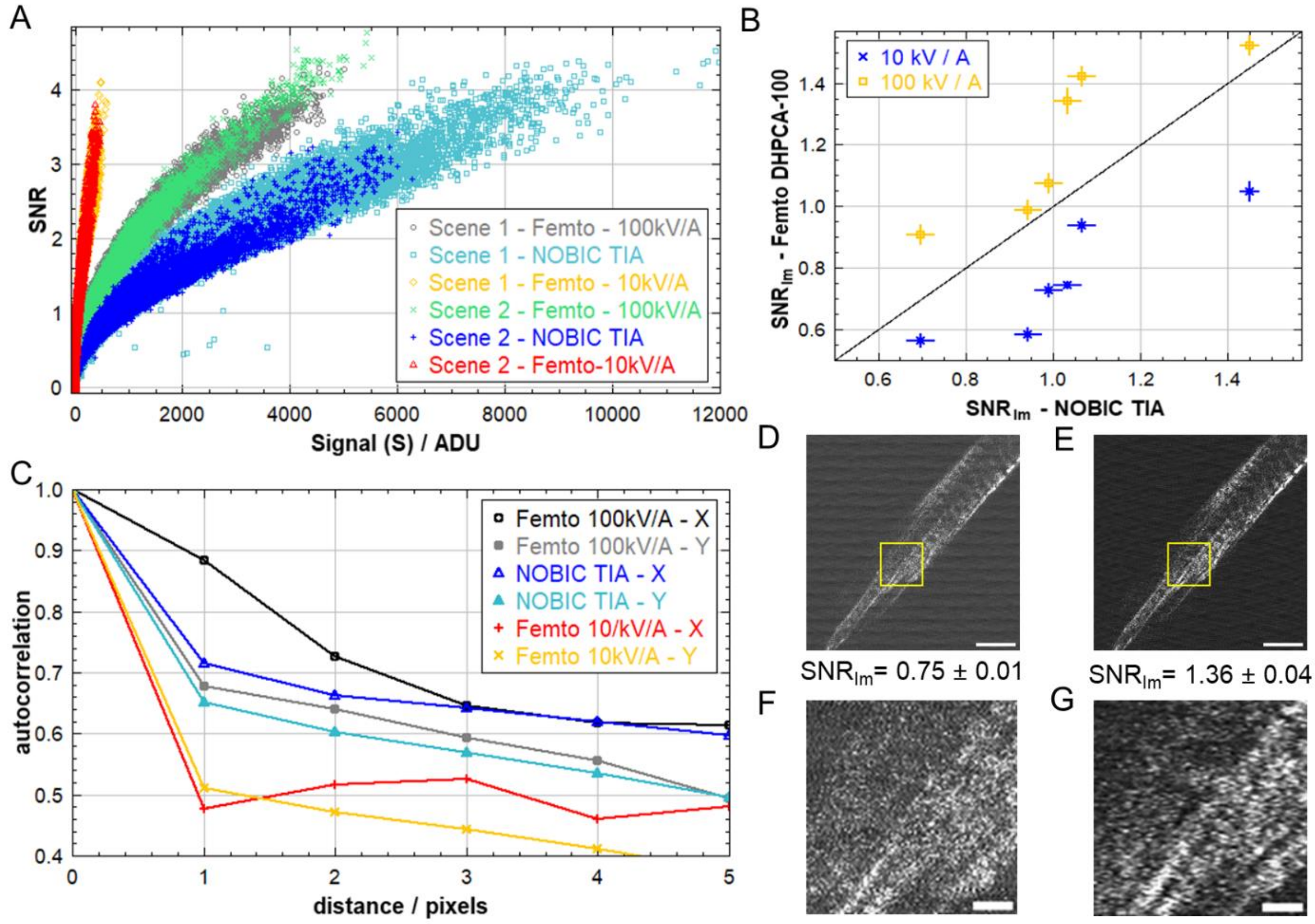

**Fig. 2:** Comparison of the NOBIC TIA (high-gain version) and Femto DHPCA-100 operated at two gain settings; (A) SNR vs. signal plot for 2 different scenes (acquired with the laser power of 19 mW and PMT gain of 650 V); (B) a plot of $SNR_{Im}$ of Femto DHPCA-100 at different gain settings vs. $SNR_{Im}$ obtained with the NOBIC TIA for the same respective scene and acquisition settings (laser power and PMT voltage). The laser power used was 13 mW or 19 mW; the PMT gain was 650 V or 700 V. The points in the plot show the mean of 3 measurements taken with the same settings at the same scene; the error bars were set as either the standard deviation from the 3 measurements or as the standard deviation from all pairs of frames in the stack whichever was higher (they were similar in all cases with the latter being higher in most). A black line through the points where the values on the horizontal and the vertical axis are equal was added to the plot for easier orientation. (C) Autocorrelation functions calculated from a single frame in the stack; an analogous plot for the other scene, data from which are shown in panel (A), is in Fig. S9. Examples of single frames acquired using Femto DHPCA-100 operated at 10 kV / A (D, F) and 100 kV / A (E, G) with the corresponding $SNR_{Im}$ (mean ± standard deviation from all pairs of frames in the stack). The scale bar in panels D and E represents 50 µm and the yellow square indicates the area shown magnified in panels F and G, in which the scale bar represents 10 µm.

*Comparison of PMTs*

We have also compared two PMT models, H7422PA-40 and H16722-40 (a later model) both from Hamamatsu Photonics (see Fig. S3). The results show that H16722-40 has a higher gain when operated at the same voltage. That means the output current of the PMT is higher for the same apparent photoelectron number $n_{p,app}$. This translates to higher output voltage of the TIA for the same $n_{p,app}$. As can be seen in Fig. S3H, H16722-40 saturates the digitiser at smaller $n_{p,app}$ than H7422PA-40. A lower-gain TIA would be, therefore, better suited for this PMT model. For comparison, a nearly perfect agreement in variance vs. signal plots was observed between two units of H7422PA-40 (see Fig. S10). Despite the differences in the gain, we found no difference between the two PMT models in SNR performance.

*Comparison with commercial 2-photon microscopes*

We have compared NOBIC 2PM fitted with the high-gain version of the NOBIC TIA with two commercial 2-photon microscopes. The same excitation intensity was used in all three cases and the dispersion pre-compensation was optimised to maximise the signal with our sample. Nikon A1R-MP permitted the most direct comparison to NOBIC 2PM as it was equipped with the identical fs-laser which we used in the custom system for this comparison, thus reducing the potential effect of laser pulse shape differences [11]. Since each microscope was equipped with a different water immersion objective lens of slightly different numerical aperture (NA), we have adjusted the laser power to account for the differences in the beam waist area and maintain the same excitation intensity. Table 3 summarises the objective lenses used at the individual microscopes, the laser power used in each case as well as the theoretical factor by which the SNR is expected to be higher than that of the lowest NA (1.0) lens, assuming fluorescence emission from the sample is isotropic and the number of detected photons proportional to the solid angle from which the lens collects light. Table 2 also includes information on the pixel integration time $\Delta t$, where known. Although each of the three microscopes was equipped with a resonant scanner operating at the same frequency (8 MHz) and the acquired images had the same number of pixels (512 x 512), the exact pixel integration time depends also on the extent to which the beam path extends beyond the imaged area, which may differ for individual microscopes. For Nikon A1R-MP the exact pixel dwell time was not shown in the software; however, we may assume it was comparable to the other two systems.

To ensure the best noise performance of the PMT, we have in each case used the highest possible PMT gain at which the image was only moderately saturated. As evidenced by Fig. S4, $SNR_{Im}$ is not influenced by moderate saturation levels and decreases with progressing saturation.

| Microscope | Objective | Objective magnification / NA | Laser power / mW | SNR factor | $\Delta t$ / ns |
|---|---|---|---|---|---|
| NOBIC 2PM | Carl Zeiss W Plan-Apochromat (Item no.: 421452-9900) | 20x / 1.00 | 9.9 | 1.00 | 88 |
| Nikon A1R-MP | Nikon CFI75 Apo 25XC W | 25x / 1.10 | 8.2 | 1.14 | - |
| Olympus FVMPE-RS | Olympus LPLN25XSVMP2 | 25x / 1.05 | 9.0 | 1.06 | 67 |

**Table 3**: Comparison of selected parameters of the compared microscopes; refer to the text for explanation.

Fig. 3 A shows the plots of SNR vs. signal level ($S$) and images of several scenes taken by each microscope with the associated $SNR_{Im}$ values. For easier comparison the $S$ values for the commercial microscopes were multiplied by a factor of 8 to account for the differences in the image bit depth compared to NOBIC 2PM (12 bit for the commercial systems and 16 bit signed for NOBIC 2PM). We found that the Nikon A1R-MP microscope achieved the highest SNR. Factors contributing to this likely include the highest NA lens from the tested microscopes and, probably more importantly, averaging of neighbouring pixels along the fast scan axis (see Fig. 3 B). This is similar to what we observed with Hamamatsu C12419 used with the NOBIC 2PM (see Fig. 1) and which can account for an increase in the SNR by a factor of almost 2 (based on fitting the initial 4 points of the autocorrelation function along the $X$ axis with a Gaussian function). It is likely caused by a TIA of insufficient bandwidth used in the Nikon A1R-MP microscope.

We have also observed that the pixel value histograms of images from the Nikon A1R-MP appear cropped on the low value side compared to those of images from the NOBIC 2PM (see Fig S11). We have tested the influence of cropping the histogram of images taken with NOBIC 2PM on the SNR and found that the SNR does not change at first and eventually decreases as the growing threshold values erode the image information content (see Fig S11). As how large portion of the histogram is clipped in the Nikon A1R-MP images cannot be ascertained, it is impossible to correctly estimate the effect of this clipping on the SNR.

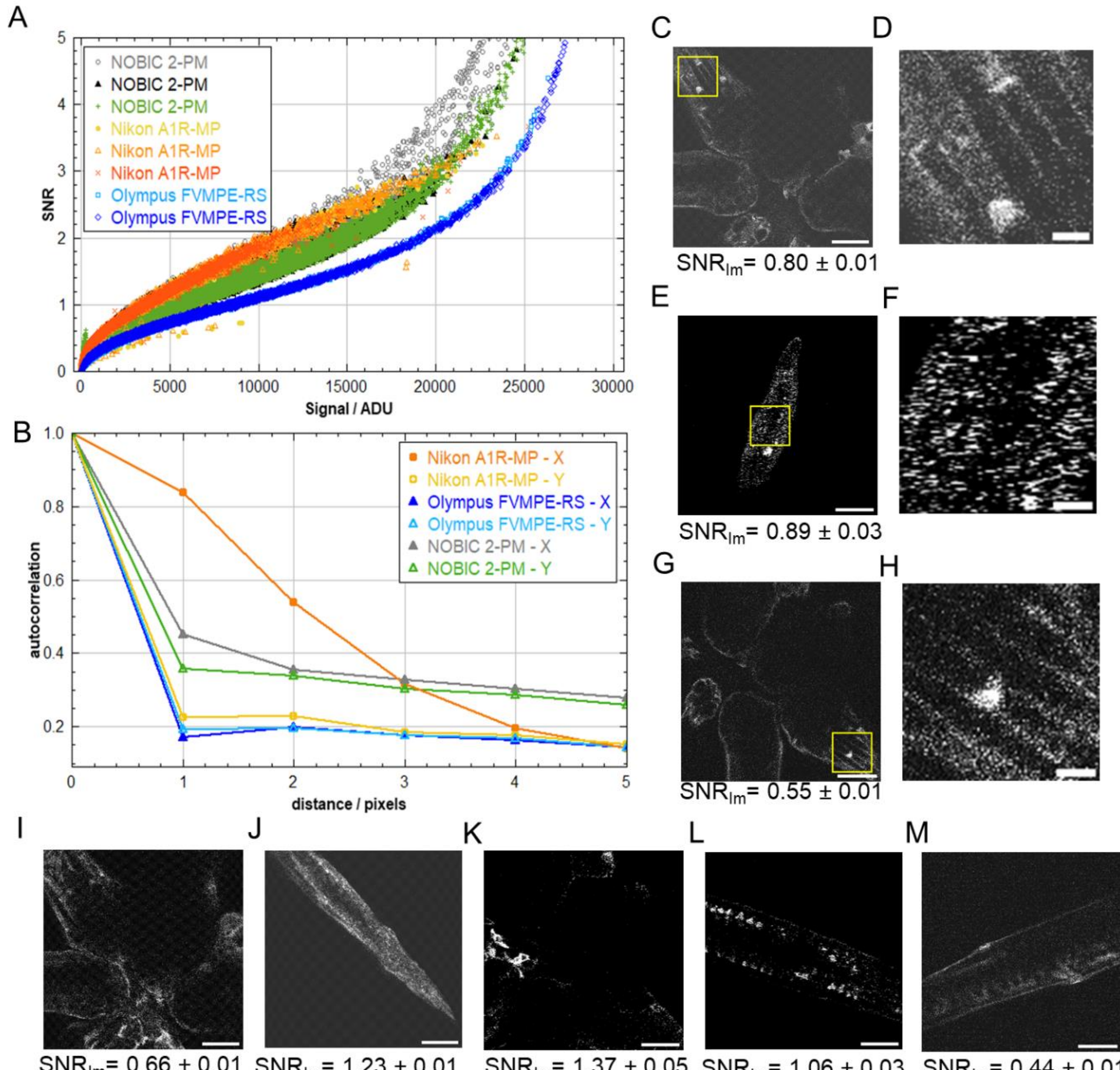

**Fig. 3:** Comparison of NOBIC 2-PM with two commercial systems: Nikon A1R-MP and Olympus FVMPE-RS. The SNR vs. signal plot showing results from two to three scenes for each microscope (A); autocorrelation functions calculated from a single frame taken from a stack acquired with each microscope (B). Examples of single frames acquired using NOBIC 2-PM (C, D, I, J), Nikon A1R-MP (E, F, K, L) and Olympus FVMPE-RS (G, H, M) with the corresponding $SNR_{Im}$ (mean ± standard deviation from all pairs of frames in the stack). The scale bar in panels C, E, G and I – M represents 50 µm and the yellow square in panels C, E and G indicates the area shown magnified in panels D, F and H, respectively, in which the scale bar represents 10 µm.

Our experiments show that Olympus FVMPE-RS has the lowest SNR. Factors contributing to this likely include the shorter pixel integration time compared to NOBIC 2PM; increasing $\Delta t$ to the same value of 88 ns would increase the SNR by a factor of 1.15. We can notice from Fig. 3 G, H that the images from this microscope show no sign of pixel averaging at all, which also contributes to the lower observed SNR, similarly to the case of Femto DHPCA-100 at 10 kV/A gain. Lastly, we can see from Fig 3 A that the level of saturation in the images taken with NOBIC 2PM and Olympus FVMPE-RS is considerably higher than in the images from Nikon

A1R-MP. This according to Fig S4 can lead to reduced SNR. In the case of Olympus FVMPE-RS we observed higher $SNR_{Im}$ at the PMT voltage used to acquire the images for comparison than at lower voltages where less or no saturation occurred (see Fig. S12). This suggests that we may not have reached the optimal PMT voltage before significant saturation in the images occurred and that higher SNR could have been possibly achieved with lower TIA gain.

**Materials and methods**

*Sample preparation*

A small piece (approximately 2 cm long) of a branch of *Asparagus setaceus* freshly cut from a plant grown indoors in a pot was attached by a piece of double-sided adhesive tape to a microscope slide. The sample was rinsed a few times with deionised water before microscope imaging. We found this treatment to reduce the intrinsic hydrophobicity of the sample and improve its wetting by deionised water used as immersion liquid during imaging. The same sample was used for experiments over a period of several weeks without any perceptible signs of degradation.

*NOBIC TIAs*

Two versions of a custom 2-channel TIA differing in their gain were built using operational amplifier LTC6269 (Analog Devices, MA). The gain of each TIA was characterized using a Tektronix MSO24 oscilloscope (Tektronix, OR). A 2.0 V peak-to-peak voltage sine wave was applied through a measured 19.610 kΩ series resistor to the TIA input, generating an input current of 101.99 μA. By sweeping the frequencies from 1 kHz to 50 MHz, the root mean square output voltage ($V_{out,rms}$) was measured at each frequency, and the gain (in dB) was calculated as $20 \cdot \log_{10}( V_{out,rms} / 101.99\ \mu A)$. A gain-versus-frequency plot was generated, and the frequency bandwidth was determined as the point where the gain dropped by 3 dB from its maximum. The results are shown in Table 1.

*NOBIC 2PM*

A custom upright 2-photon microscope, described earlier [11], was built around an Axio Examiner (Carl Zeiss, Germany) frame. SPARK Alcor Duo (SPARK Lasers, France) fs-laser providing laser lines of 920 nm and 1064 nm was used as the excitation source unless stated otherwise. The laser beam was expanded two times by a reflective beam expander (#37-193, Canopus Reflective Beam Expander, Edmund Optics, NJ) and scanned by a custom scan engine containing an 8 kHz resonant scanner (CRS8K, Cambridge Technology, MA) for the fast scan axis and a non-resonant galvanometer scanner (Saturn 5B, ScannerMAX, FL) for the slow scan axis. The scanned beam then passed through a custom-designed scan and tube lens in a $4f$ configuration and passed the main dichroic mirror (BLP01-785R, Semrock, NY) and overfilled the back aperture of a water dipping objective W Plan-Apochromat (Item no.: 421452-9900, Carl Zeiss, Germany) mounted on a piezo scanner (ND72Z2LAQ PIFOC, Physik Instrumente, Germany).

The fluorescence signal collected by the objective was reflected by the main dichroic mirror, passed through a laser blocking filter (ET750sp-2p8, Chroma, VT) and was separated by a secondary dichroic mirror (FF562-Di03, Semrock, NY) into two beams which after passing through emission filters, either FF01-520/70 (green channel) or FF01-593/46 (red channel) (both from Semrock, NY), unless stated otherwise, were detected by two cooled PMTs.

Identical PMTs (H7422PA-40, Hamamatsu, Japan) were used for both channels unless stated otherwise. The PMTs were mounted on a liquid-cooled heatsink maintained at 20°C. The output of the detectors was amplified by a TIA, the high gain version of NOBIC TIA was used unless stated otherwise, and digitised by a data acquisition card (vDAQ, MBF Bioscience, VT). The whole system was controlled by ScanImage software (MBF Bioscience, VT).

*Data acquisition*

After finding a suitable region of the sample based on epifluorescence observation through the eyepiece, a stack of 500 frames was acquired. The image format was 512 x 512 pixels unless stated otherwise. The laser power at the sample plane was measured by PM100D power meter equipped with S170C probe (both Thorlabs, Germany) and laser dispersion compensation was adjusted to maximise the signal at the given laser power. The acquired image stacks were analysed in FIJI [12,13] using custom macros to generate the variance vs. signal and the SNR vs. signal plots and calculate correlations, $SNR_{Im}$ and $SNR_{Corr}$. The macros will be made publicly available through our GitHub repository (https://github.com/NTU-NOBIC/FIJI-Macros) latest by the date of publication of this study.

*Comparison PMTs and TIAs*

To compare two PMT models, H7422PA-40 and H16722-40 (both Hamamatsu Photonics, Japan), the secondary dichroic was replaced with a 50/50 beam splitter (BSW, Thorlabs) and identical emission filters were installed in front of both PMTs (FF01-520/70, Semrock, NY). 920 nm laser line was used for excitation. Two combinations were each PMT model was installed at the first and the second position, respectively, were tested.

The same configuration was used for comparing the high-gain version of the custom TIA with C12419 TIA (Hamamatsu Photonics, Japan), only in this case 2 units of H7422PA-40 PMT were used and the output from each was connected to a different TIA. Both combinations were tested, output from the first PMT connected to C12419 and from the second PMT to the NOBIC TIA (example data shown in Fig. 1) and vice versa (example data shown in Fig. S6).

For the comparison between the two versions of the NOBIC TIA and between Femto DHPCA-100 (FEMTO Messtechnik, Germany) and the NOBIC TIA, we used the same H7422PA-40 PMT unit at the same position and did the measurements with the different TIAs in sequence. As we did not expect the differences to be large, we wanted to avoid potential confounding effects from imperfect splitting of the signal between the two PMTs and/or differences in performance between individual PMT units. Care was taken to maintain the same focus position between the sequential measurements. For both comparisons the standard configuration with a secondary dichroic and a green and a red channel was used, 1064 nm laser line (12.7, 19 or 29 mW power) and the H7422PA-40 PMT at the second position (red channel) operated at 650 V or 700 V gain.

*Comparison with commercial 2-photon microscopes*

Measurements for the comparison with the commercial microscopes were done with FemtoFiber ultra 920 laser (Toptica, Germany) as the excitation light source; 9.9 mW power at the sample was used and signal from the green channel PMT was collected. The results were compared with those from Nikon A1R-MP (Nikon, Japan) equipped with a FemtoFiber ultra 920 laser (Toptica, Germany), CFI75 Apo 25XC W water dipping objective (Nikon, Japan) and an emission filter for GFP (the exact spectral band not known) in front of a GaAsP PMT and from Olympus FVMPE-RS (Olympus, Japan) equipped with a Chameleon Vision laser (Coherent, UK), which was tuned to 920 nm, LPLN25XSVMP2 water dipping objective (Olympus, Japan) and 495-540 nm bandpass emission filter in front of a GaAsP PMT. Laser power used was adjusted according to the objective NA to achieve excitation intensity comparable to that used with NOBIC 2PM; the actual laser power values are in Table 3.

## Conclusions

We have investigated the noise performance of a custom built and two commercial 2-photon microscopes including a comparison of three TIAs used in conjunction with the custom microscope. We used both pixel-wise and image-based metrics of the SNR. Although the results of the two approaches may differ, in most situations including those encountered in this study, they follow the same trends and provide comparable values. The differences in noise performance we observed in the comparison of the NOBIC TIAs with commercial ones can be mostly attributed to differences in frequency bandwidth of the TIAs. While we cannot entirely exclude other contributing factors, the averaging of pixel values caused by insufficient frequency bandwidth of some of the TIAs could alone explain the observed trends in the SNR.

We have also compared the noise performance of the NOBIC 2-PM fitted with the NOBIC TIA with two commercial 2-photon microscopes. As in this case there are many more factors influencing the results (laser pulse shape, light collection efficiency, emission filters, detector quantum efficiency, etc.), the individual effects of which may not be feasible to isolate, we did not attempt a fully quantitative comparison. Overall, we found the NOBIC 2-photon microscope SNR compare favourably to the commercial systems. One of the commercial systems possibly slightly overperformed our custom system in the SNR; however, the high SNR was in this case achieved at the cost of compromised spatial resolution, likely because of a limited TIA frequency bandwidth. Too high TIA gain was possibly a factor contributing to the lower SNR of the other commercial microscope. Significant levels of saturation prevented further increase of PMT voltage, which means that we may not have operated the PMT at a voltage where it gives the best SNR.

Although the contribution of the TIA to the microscope noise may not be as significant as that of the PMT, an appropriate choice of the TIA parameters is necessary to reach the best possible SNR. While insufficient frequency bandwidth may improve the SNR, it happens at the cost of spatial resolution. Although spatial resolution may not be the highest priority in applications where fast scanning is typically used, we would still consider a sufficient frequency bandwidth preferable as equivalent SNR enhancement can be achieved through pixel size setting or through pixel averaging as post processing, without losing the ability to acquire images at full resolution when needed.

Lastly, we believe that the measurements presented here can be useful for those interested in characterising and comparing SNR performance and linearity range of their microscopes and those who wish to include SNR tests to regular microscope performance monitoring alongside point spread function or field uniformity measurements [7,14].

## Acknowledgements

The assistance of Dr. Ruey Kuang and Dr. Ma Xiaoxiao with the access to and measurements at the commercial 2-photon microscopes is gratefully acknowledged.

**Competing interests**

Authors declare no competing interests.

**Funding declaration**

The research was funded by Nanyang Technological University, Singapore and Lee Kong Chian School of Medicine, Nanyang Technological University, Singapore.

**References**

1. J. B. Pawley, "Fundamental Limits in Confocal Microscopy BT - Handbook Of Biological Confocal Microscopy," in J. B. Pawley, ed. (Springer US, 2006), pp. 20–42.

2. J. B. Pawley, "Points, Pixels, and Gray Levels: Digitizing Image Data BT - Handbook Of Biological Confocal Microscopy," in J. B. Pawley, ed. (Springer US, 2006), pp. 59–79.

3. C. J. R. Sheppard, X. Gan, M. Gu, and M. Roy, "Signal-to-Noise Ratio in Confocal Microscopes BT - Handbook Of Biological Confocal Microscopy," in J. B. Pawley, ed. (Springer US, 2006), pp. 442–452.

4. R. Diekmann, J. Deschamps, Y. Li, T. Deguchi, A. Tschanz, M. Kahnwald, U. Matti, and J. Ries, "Photon-free (s)CMOS camera characterization for artifact reduction in high- and super-resolution microscopy," Nat. Commun. **13**, (2022).

5. J. Art, "Photon Detectors for Confocal Microscopy BT - Handbook Of Biological Confocal Microscopy," in J. B. Pawley, ed. (Springer US, 2006), pp. 251–264.

6. N. Matsunaga, ed., *Photomultiplier Tubes - Basics and Applications*, 4th ed. (Hamamatsu K.K., 2017).

7. A. Ferrand, K. D. Schleicher, N. Ehrenfeuchter, W. Heusermann, and O. Biehlmaier, "Using the NoiSee workflow to measure signal-to-noise ratios of confocal microscopes," Sci. Rep. **9**, 1165 (2019).

8. R. Heintzmann, P. K. Relich, R. P. J. Nieuwenhuizen, K. A. Lidke, and B. Rieger, "Calibrating photon counts from a single image," (2016).

9. D. C. Y. Ong and K. S. Sim, "Single Image Signal-to-Noise Ratio (SNR) Estimation Techniques for Scanning Electron Microscope - A Review," IEEE Access **12**, 155747–155772 (2024).

10. J. Frank and L. Al-Ali, "Signal-to-noise ratio of electron micrographs obtained by cross correlation," Nature **256**, 376–379 (1975).

11. S. P. Chong and P. Török, "Influence of laser pulse shape and cleanliness on two-photon microscopy," Opt. Contin. **3**, 552–564 (2024).

12. J. Schindelin, I. Arganda-Carreras, E. Frise, V. Kaynig, M. Longair, T. Pietzsch, S. Preibisch, C. Rueden, S. Saalfeld, B. Schmid, J.-Y. Tinevez, D. J. White, V. Hartenstein, K. Eliceiri, P. Tomancak, and A. Cardona, "Fiji: an open-source platform for biological-image analysis," Nat. Methods **9**, 676–682 (2012).

13. C. A. Schneider, W. S. Rasband, and K. W. Eliceiri, "NIH Image to ImageJ: 25 years of image analysis," Nat. Methods **9**, 671–675 (2012).

14. J. Jonkman, C. M. Brown, G. D. Wright, K. I. Anderson, and A. J. North, "Tutorial: guidance for quantitative confocal microscopy," Nat. Protoc. **15**, 1585–1611 (2020).

# Supporting material for:

# Characterisation of the signal to noise ratio of 2-photon microscopes

by

*Radek Macháň, Shau Poh Chong, Khee Leong Lee & Peter Török*

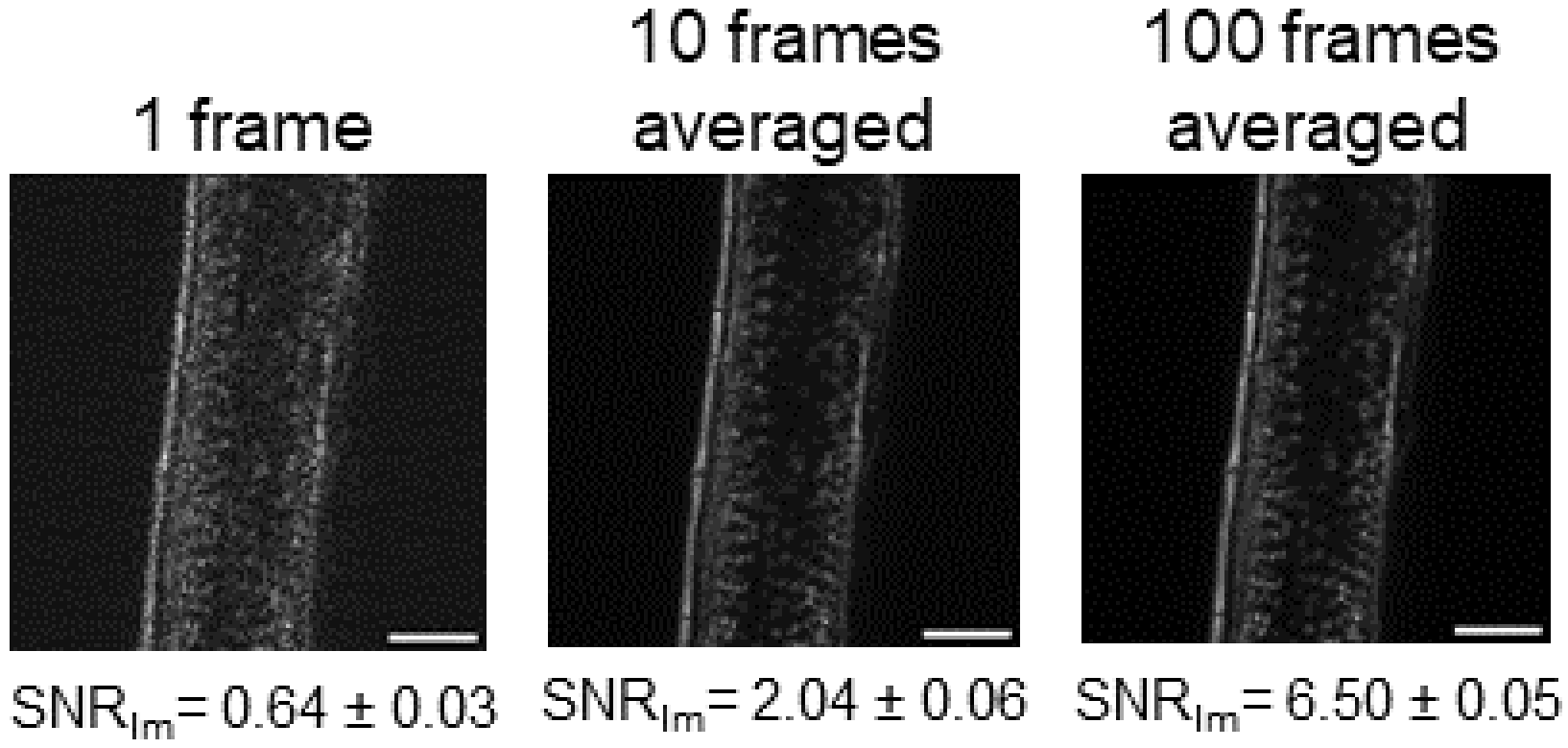

**Fig. S1:** An example of the influence of frame averaging on SNR of a two photon microscope image. A single frame and averages of 10 and 100 frames from the same stack of 500 frames are shown together with the corresponding $SNR_{Im}$ calculated according to equation (5). The $SNR_{Im}$ values are shown as the mean and standard deviation calculated from all pairs of consecutive images in the stack; for a single frame there are 499 pairs, for the average of 10 frames 49 pairs and for the average of 100 frames 4 pairs. The scale bar represents 50 μm.

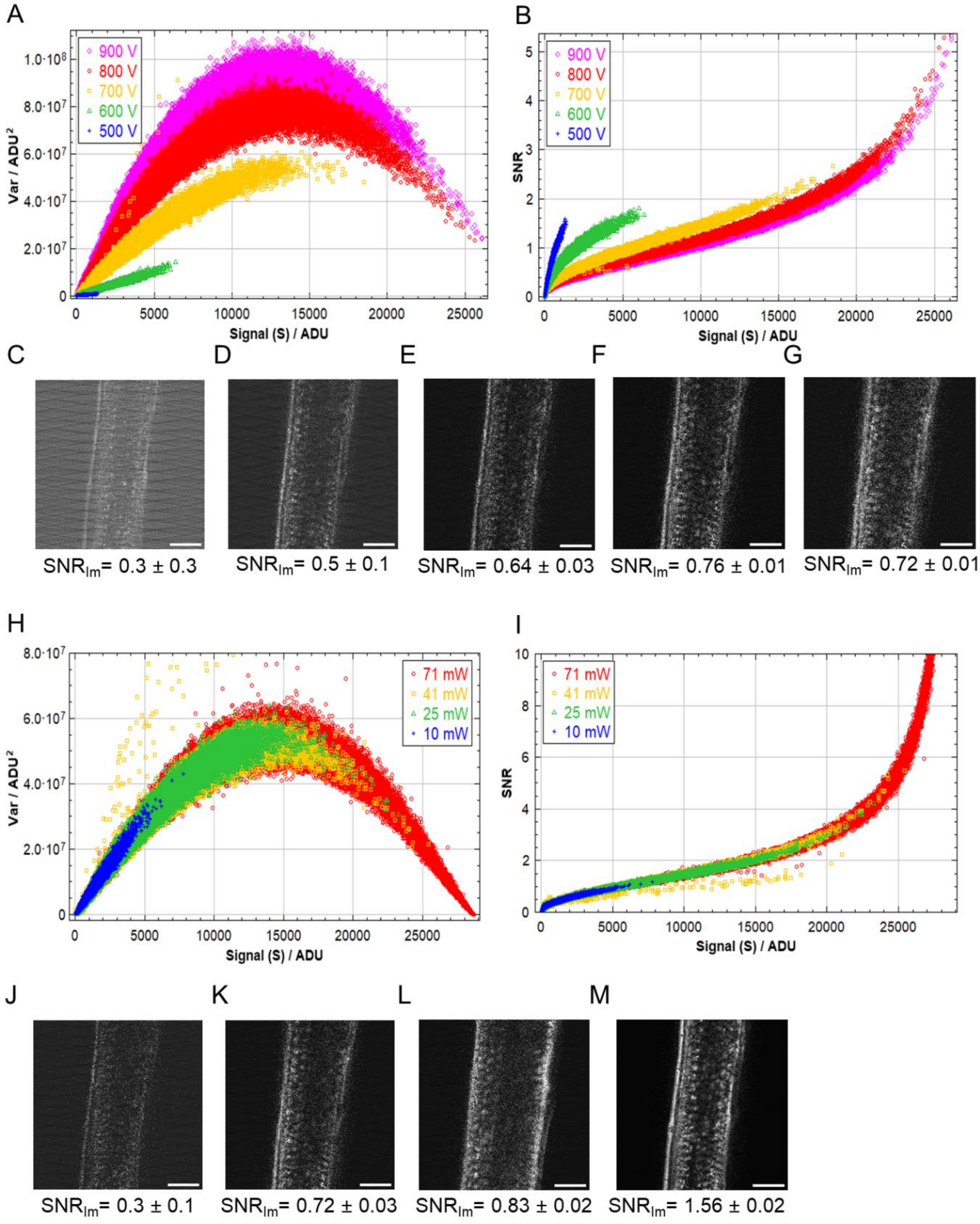

**Fig. S2:** Variance vs. signal plots (A, H) and corresponding SNR vs. signal plots (B, I) for different imaging scenarios; in the first case (A, B) the excitation power was constant (25 mW at $\lambda = 920$ nm) and the PMT voltage (and therefore the gain $K$) was varied; in the second case (H, I) the gain was constant (700 V) and the laser power was varied. Panels C – G and J – K show single frames corresponding to the curves in the plots in panels A and B or H and I, respectively, in the order of the increasing gain or laser power, respectively. The values of $SNR_{Im}$ calculated according to the equation (5) are shown below the respective images. The scale bar represents 50 μm.

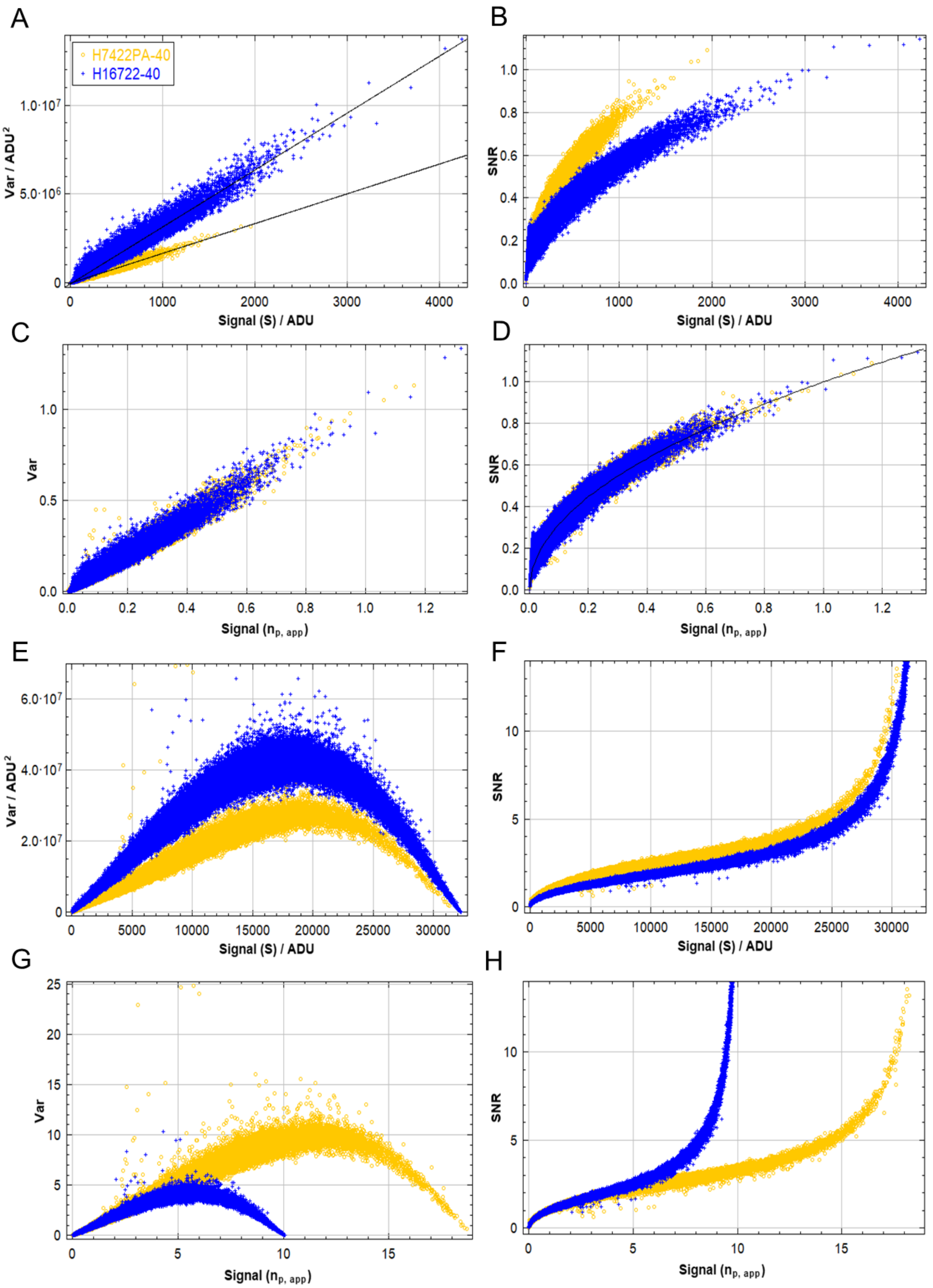

**Fig. S3:** Variance vs. signal plots (A, C, E, G) and corresponding SNR vs. signal plots (B, D, F, H) showing the comparison of two PMT models H7422PA-40 (orange circles), and H16722-

40 (a later model, blue crosses). The data were taken from images of 2 scenes acquired with different power of the 920 nm laser (A – D scene 1, 10 mW; E – H scene 2, 85 mW); the same PMT voltage (700 V) was used in all cases. The gain $K$ was determined as the slope of the linear fits to the variance vs. signal plots in the unsaturated case (A); the fits are shown in black solid lines. The values of $K$ determined thus, were then used to convert the signal from ADU (used in panels A, B, E, F) to the apparent photoelectron numbers $n_{p,app}$ (used in panels C, D, G, H). The black solid line in panel D shows $\sqrt{n_{p,app}}$.

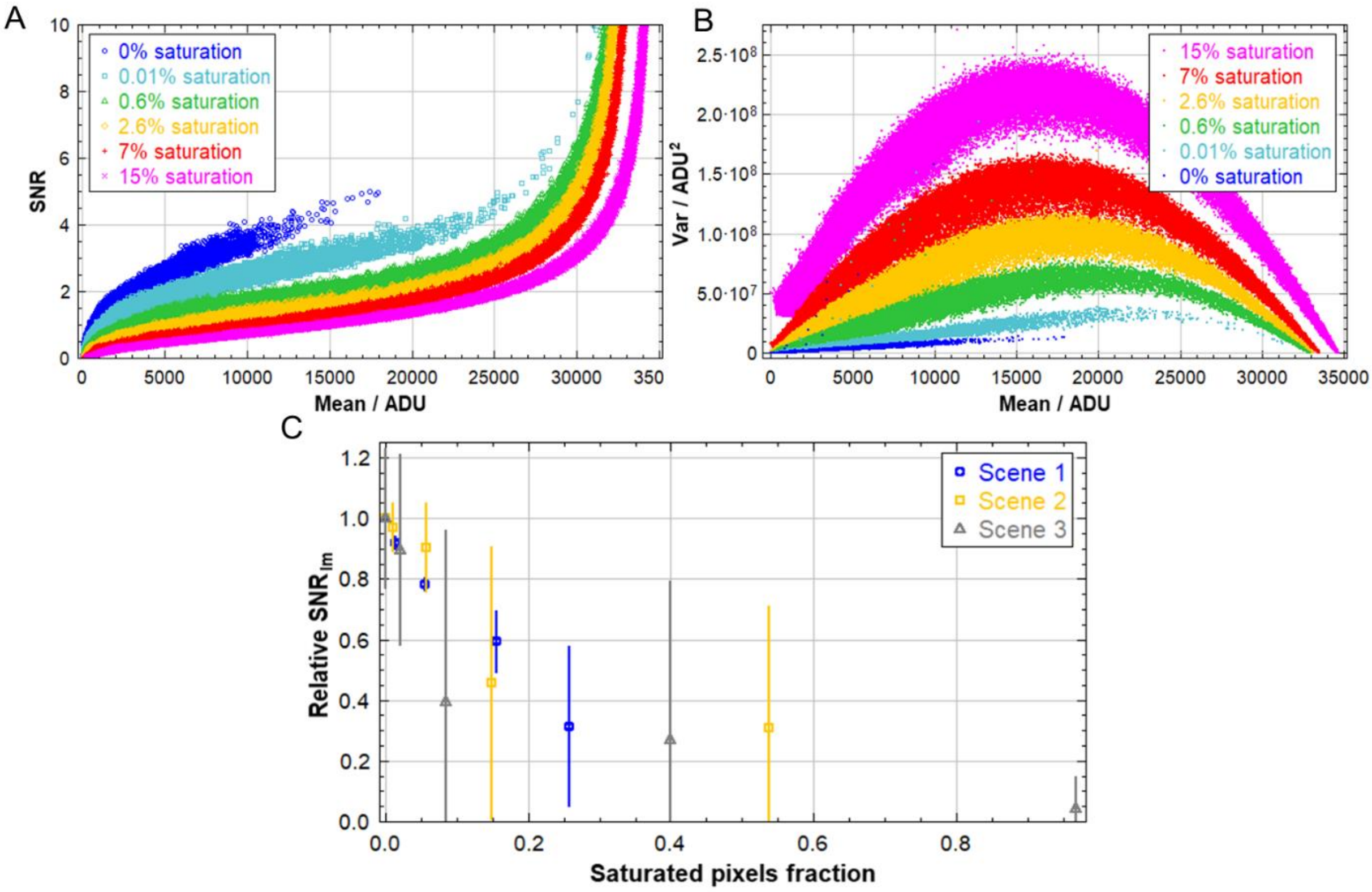

**Fig. S4:** A simulation of the influence of signal saturation on the SNR. Image stacks free from saturated pixels were multiplied by increasing constants to generate stacks with growing fractions of pixels reaching the upper limit for the image bit-depth. Examples of the SNR vs. signal plots (A) and corresponding variance vs. signal plots (B) for the different fractions of saturated pixels are shown. Panel C shows a plot of the relative $SNR_{Im}$ (mean ± standard deviation from all pairs of frames in the stack) as a function of the saturated pixels fraction for 3 different scenes acquired under different settings and differing in $SNR_{Im}$ of the original stack free from saturation (scene 1: 650 V PMT gain, 1064 nm laser at 29 mW power, $SNR_{Im}$ = 1.60 ± 0.03; scene 2: 650 V PMT gain, 1064 nm laser at 19 mW power, $SNR_{Im}$ = 1.03 ± 0.07; scene 3: 700 V PMT gain, 920 nm laser at 56 mW power, $SNR_{Im}$ = 0.5 ± 0.1). The high-gain version of NOBIC TIA was used in all 3 cases.

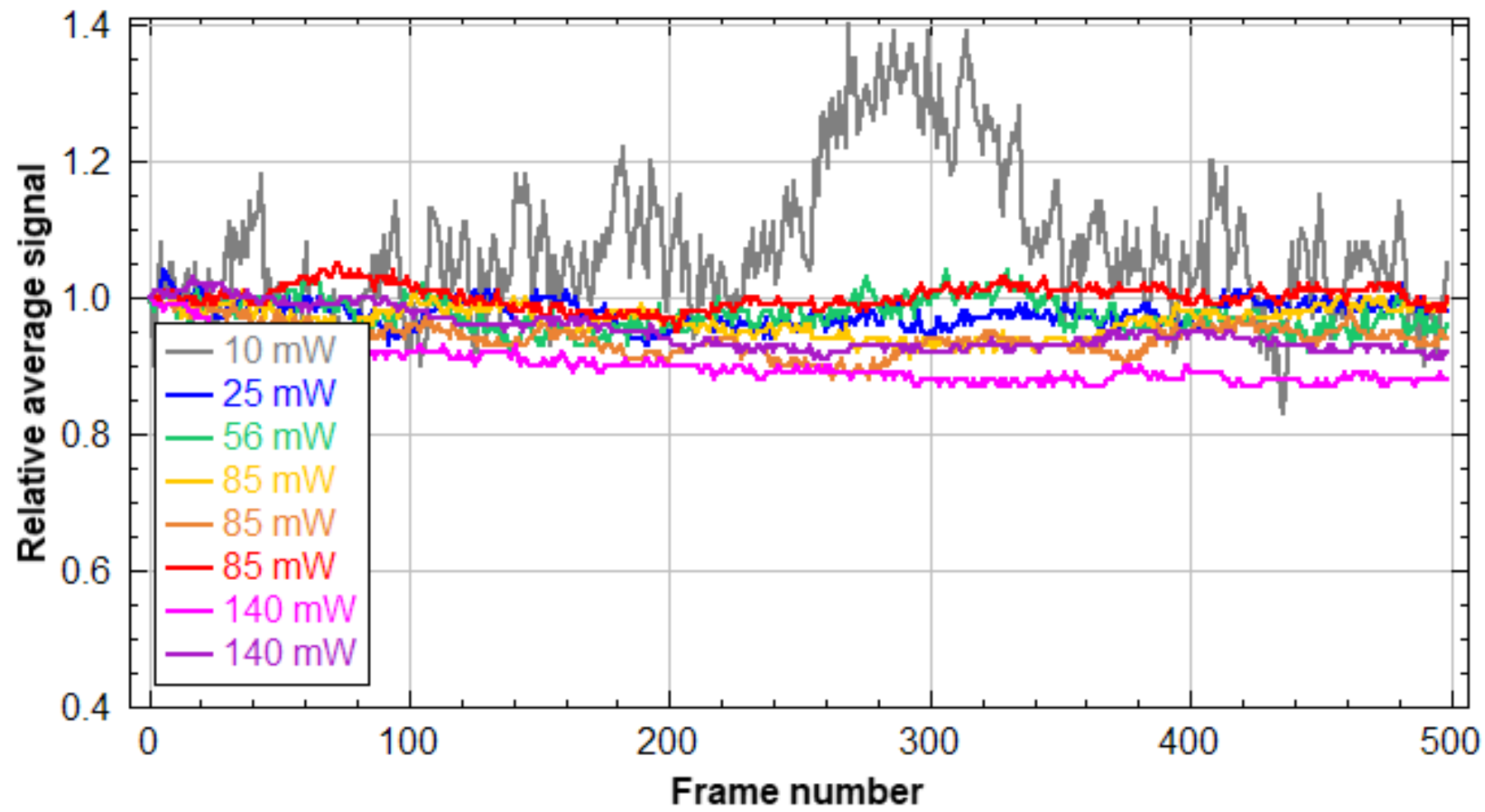

**Fig. S5:** Temporal evolution of the measured fluorescence signal during the acquisition of 500-frame stacks using different powers of the 920 nm excitation laser. The signal was plotted as the average from all pixels in each frame normalised to the average signal in the first frame for easier comparison.

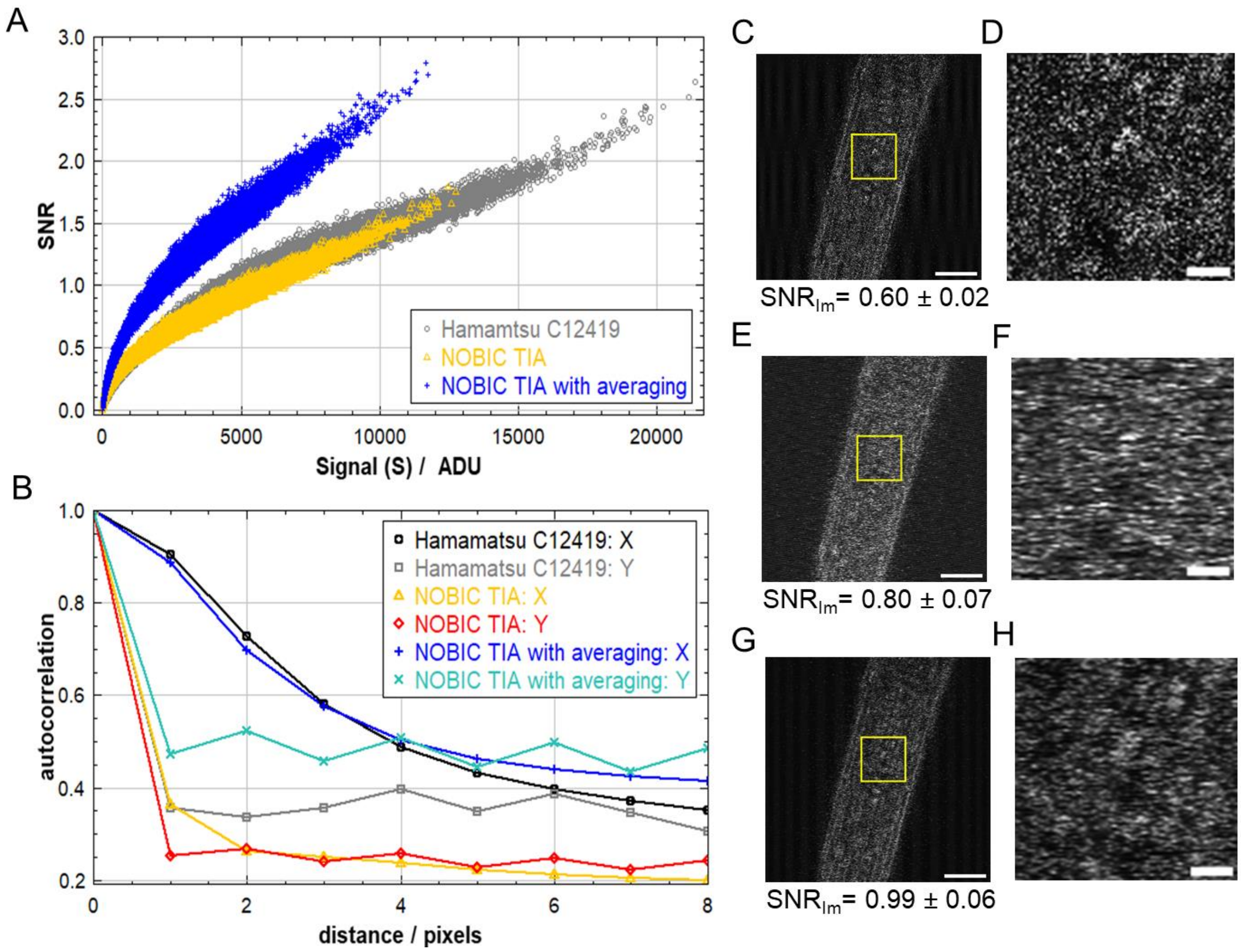

**Fig. S6:** Comparison of the NOBIC TIA (high-gain version) and Hamamatsu C124190 (a figure analogous to Fig. 1 showing data from another scene): the SNR vs. signal plot (A), autocorrelation functions calculated from a single frame in the stack (B) and examples of single

frames acquired using NOBIC TIA (C, D) and Hamamatsu C124190 (E, F), with the corresponding $SNR_{Im}$ (mean ± standard deviation from all pairs of frames in the stack). The scale bar in panels C and E represents 50 µm and the yellow square indicates the area shown magnified in panels D and F, in which the scale bar represents 10 µm. An example of a frame acquired with NOBIC TIA after averaging along $X$ axis (as described in the text) is shown together with a magnified view (G, H) and data for an image averaged in that way are plotted in panels A and B.

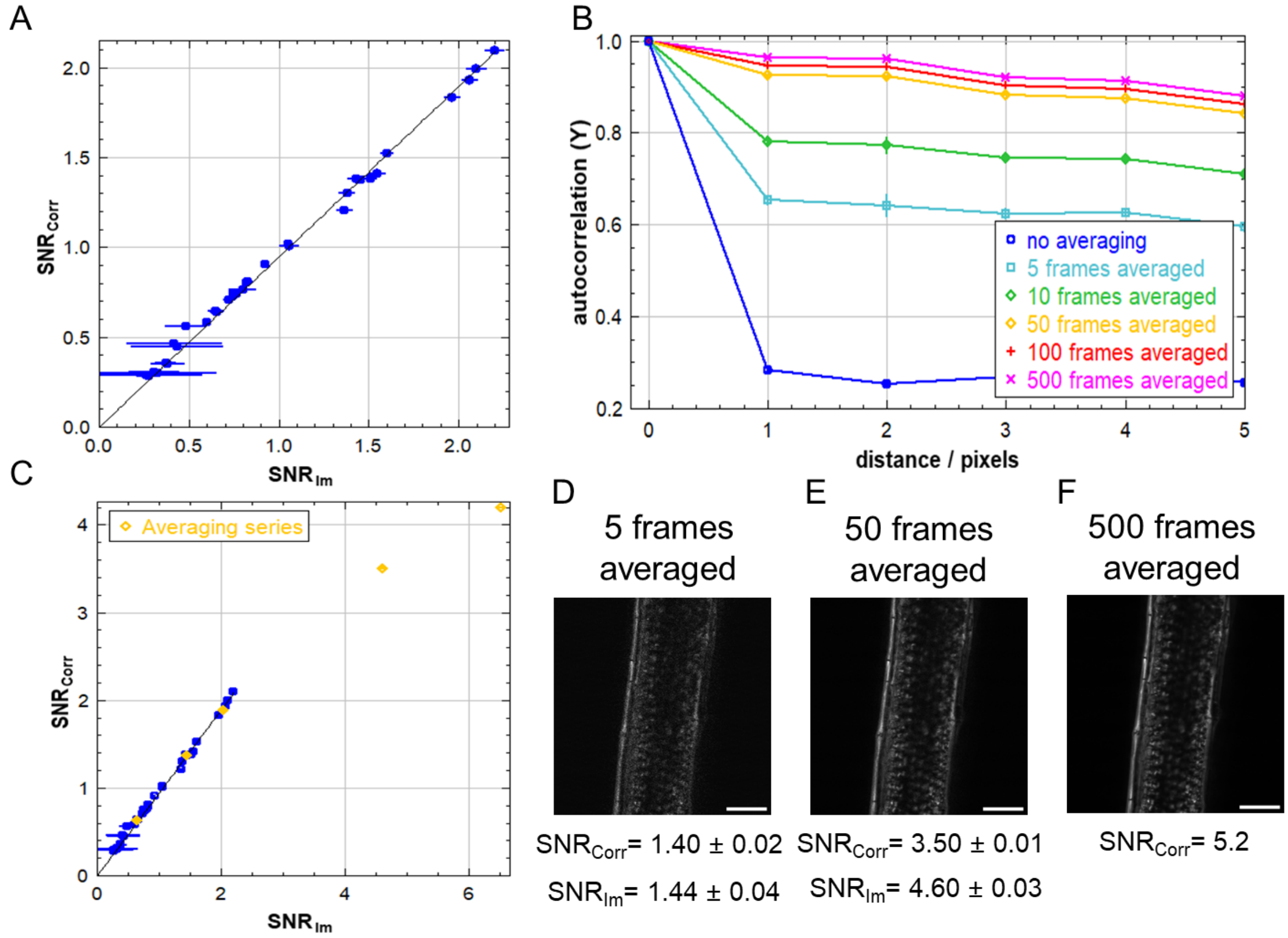

**Fig. S7:** A comparison of $SNR_{Corr}$ calculated according to equation (8) with $SNR_{Im}$ calculated according to equation (5). (A) A plot of $SNR_{Corr}$ vs. $SNR_{Im}$ of images acquired with different settings (excitation laser line 920 nm or 1064 nm, laser power in the range from 13 to 140 mW, PMT gain from 500 V to 800 V, all compared TIA models). The values are the mean calculated from all frames ($SNR_{Corr}$) or pairs of frames ($SNR_{Im}$) in the stack ± standard deviation. A linear fit going through the origin is shown by a black line; the slope is 0.95. $SNR_{Corr}$ is expected to underestimate the actual SNR by neglecting the noise-independent loss in correlation at the distance of 1 pixel. (B) Autocorrelation along $Y$ axis calculated from a stack of images in which different numbers of subsequent frames were averaged (the same stack used in Fig S1 to show the influence of the frame averaging on the SNR). (C) Same data as in panel A with the values for the frame averaging series shown in panel B added. As the SNR increases, the relative contribution of the noise to the loss in correlation at the distance of 1 pixel decreases and $SNR_{Corr}$ increasingly underestimates the actual SNR. (D, E, F) Examples of frames taken from the stacks with different numbers of averaged frames that were used to generate the data

shown in panel B. The values of $SNR_{Corr}$ and $SNR_{Im}$ (mean ± standard deviation from the stack) are shown below the respective images (except of the case of 500 frames averaged, where the whole stack was reduced to a single frame). The scale bar represents 50 μm.

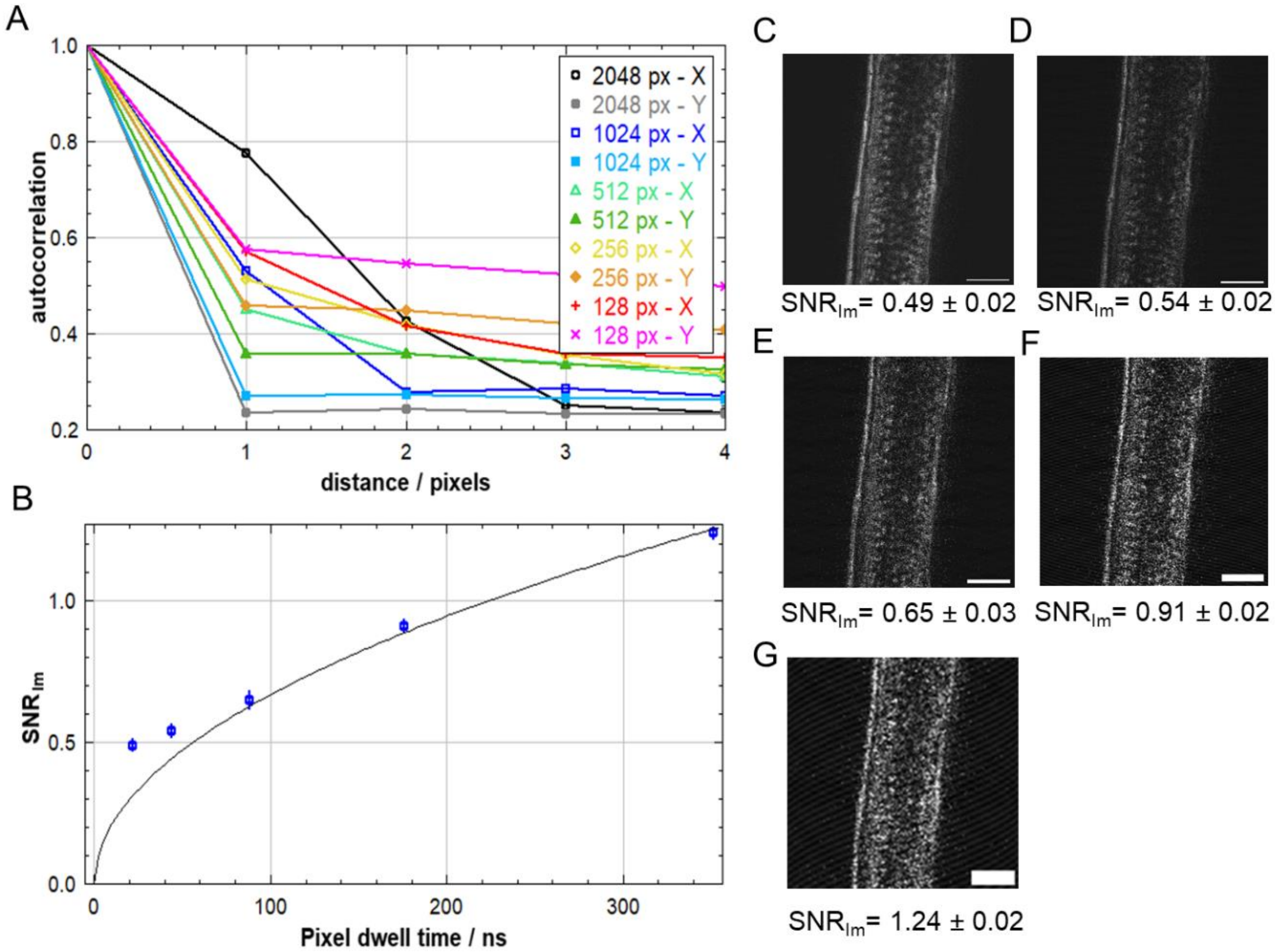

**Fig. S8:** An illustration of the influence of the pixel dwell time on pixel autocorrelations along the fast scanning (*X*) axis and on the SNR. Different pixel dwell times were achieved by changing the number of pixels per line from 2048 to 128 resulting in pixel dwell times ranging from 22 ns to 351.5 ns. (A) autocorrelation functions calculated from a single frame in the stack for stacks with different number of pixels per line. (B) Plot of $SNR_{Im}$ as a function of pixel dwell time; the error bars represent the standard deviation from all pairs of frames in the stack. The black solid line represents the expected dependence if the effect of the limited TIA bandwidth was neglected and was calculated as $SNR_{Im} = A\sqrt{\Delta t}$, where the constant $A$ was determined by fitting through the last two points (with the longest pixel dwell time) which are assumed to be only marginally affected by the TIA bandwidth limitations. (C – G) Examples of single frames acquired with different pixel dwell times increasing from 22 ns (C) to 351.5 ns (G) with the corresponding $SNR_{Im}$ (mean ± standard deviation from all pairs of frames in the stack); the scale bar in panels C – G represents 50 μm.

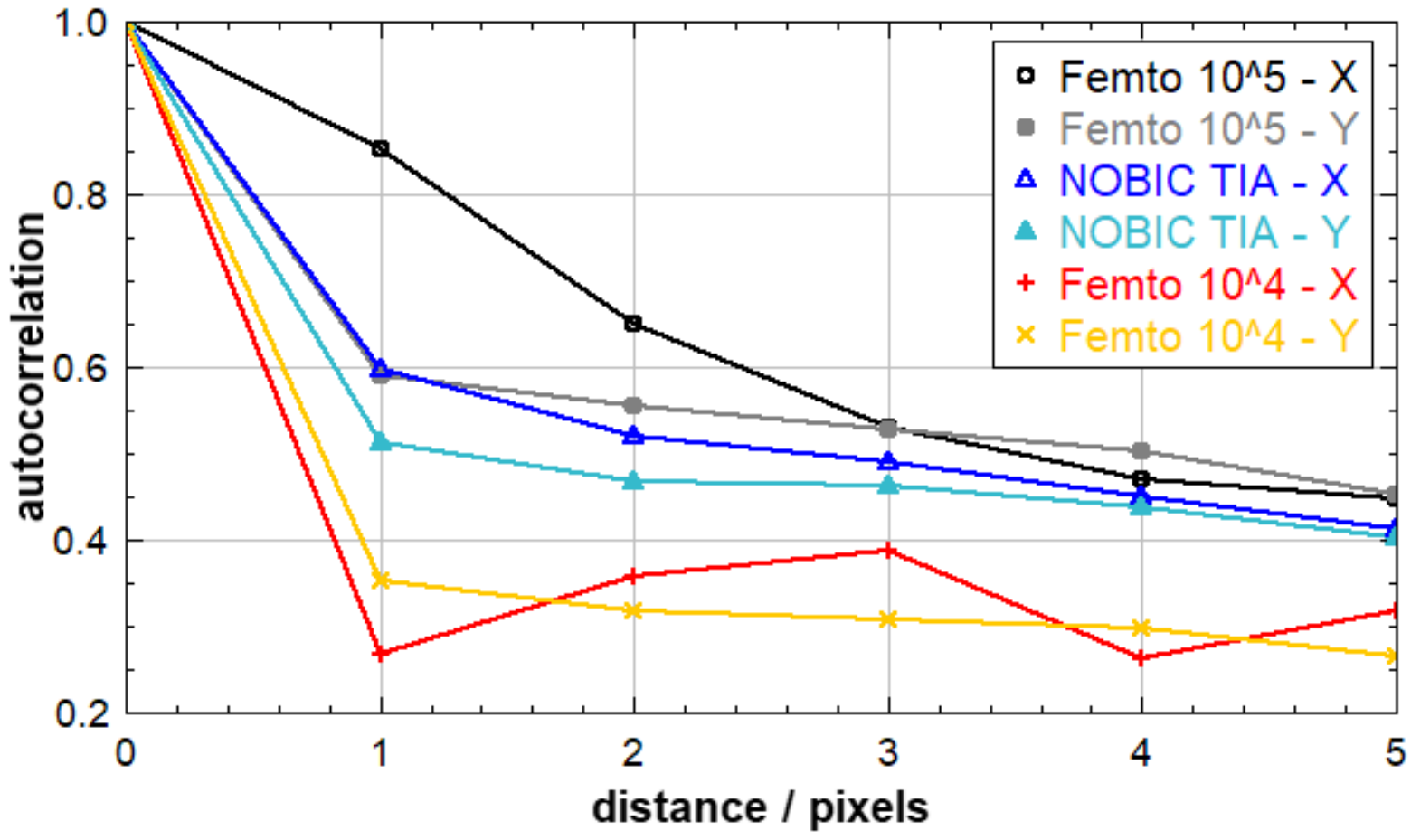

**Fig. S9:** Comparison of the NOBIC TIA (high-gain version) and Femto DHPCA-100 operated at two gain settings: autocorrelation functions calculated from a single frame in the stack; this plot is analogous to the one in Fig. 2C; refer to Fig. 2 legend for more information.

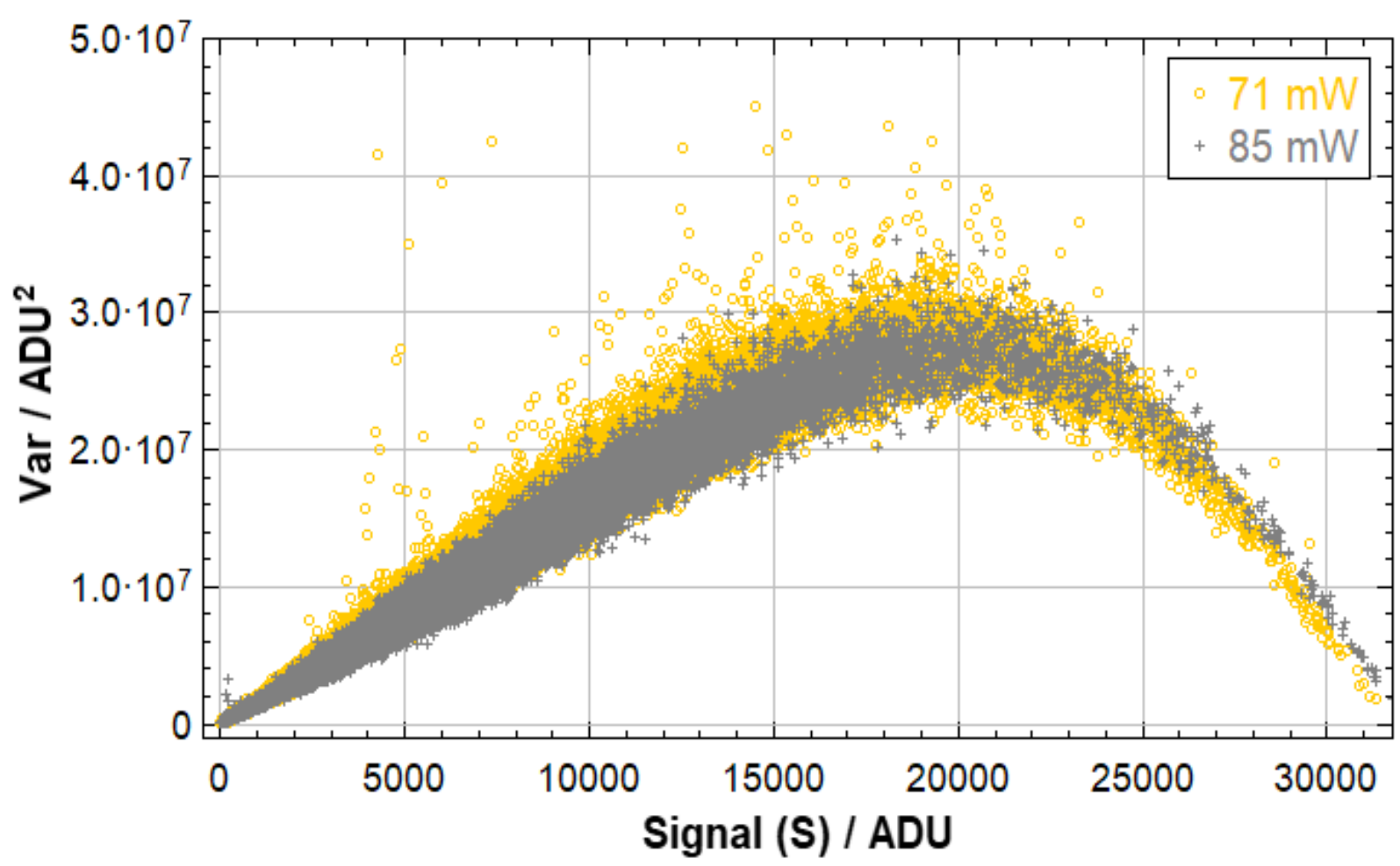

**Fig. S10:** Variance vs. signal plots comparing 2 units of H7422PA-40 PMT. The data were recorded using the same PMT gain (700 V) and the same NOBIC TIA but at different scenes and with different power of the 920 nm laser (indicated in the figure).

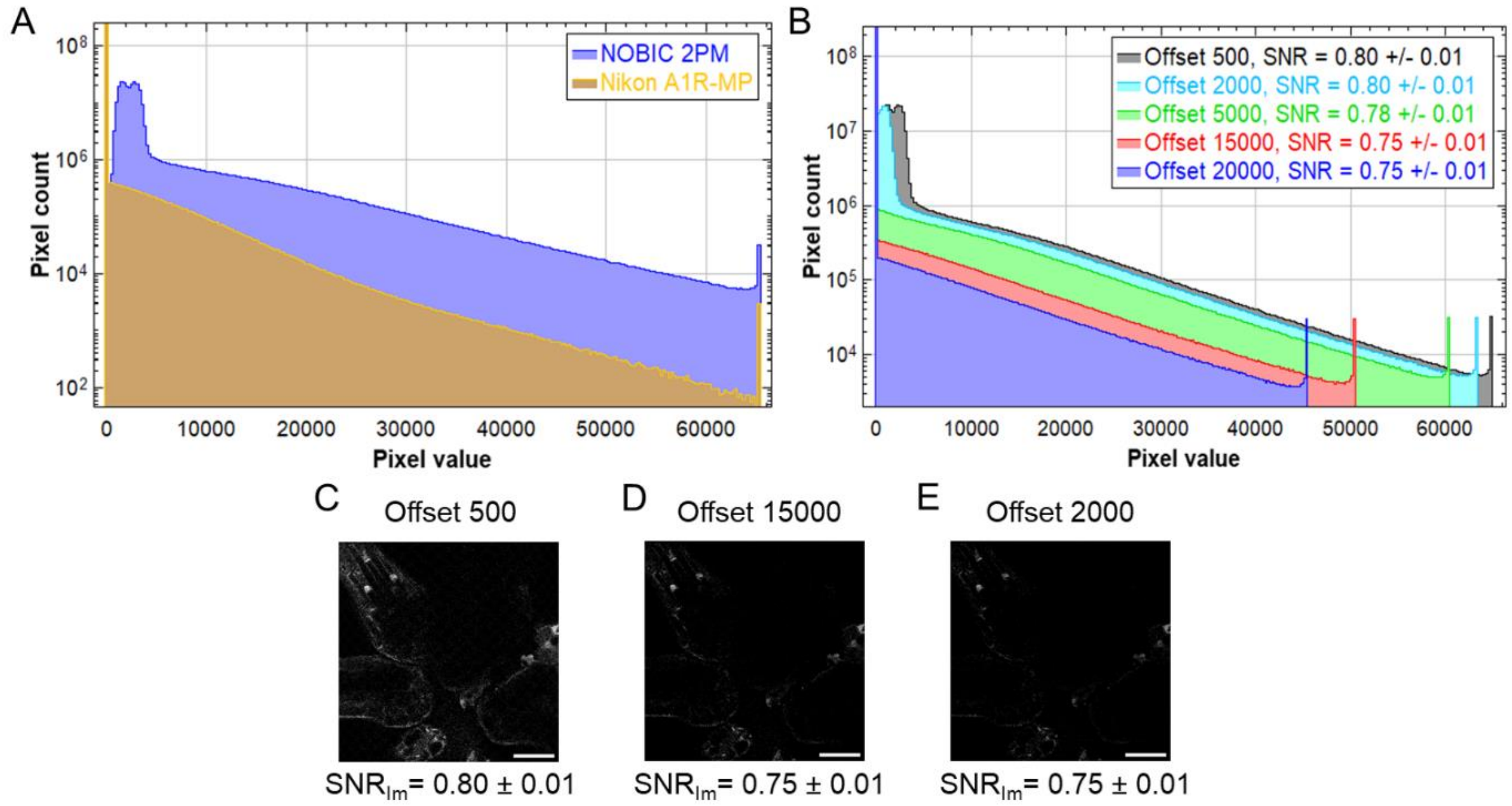

**Fig. S11:** (A) Examples of histograms of images acquired with Nikon A1R-MP and NOBIC 2PM (corresponding to images shown in Fig. 3 C and E). Because of the different bit-depth of the images recorded by the two instruments (12-bit for Nikon A1R-MP and 16-bit signed for NOBIC 2PM), both images were converted to 16-bit unsigned for easier comparison. (B) Histograms of the image acquired with NOBIC 2PM after different offsets have been subtracted from the image; $SNR_{Im}$ (mean ± standard deviation from all pairs of frames in the stack) of the resulting images is shown in the plot legend. (C, D, E) Examples of single frames taken from the stack after subtraction of different offsets (the offset value is indicated in the figure) with the corresponding $SNR_{Im}$ (mean ± standard deviation from all pairs of frames in the stack); the scale bar represents 50 μm.

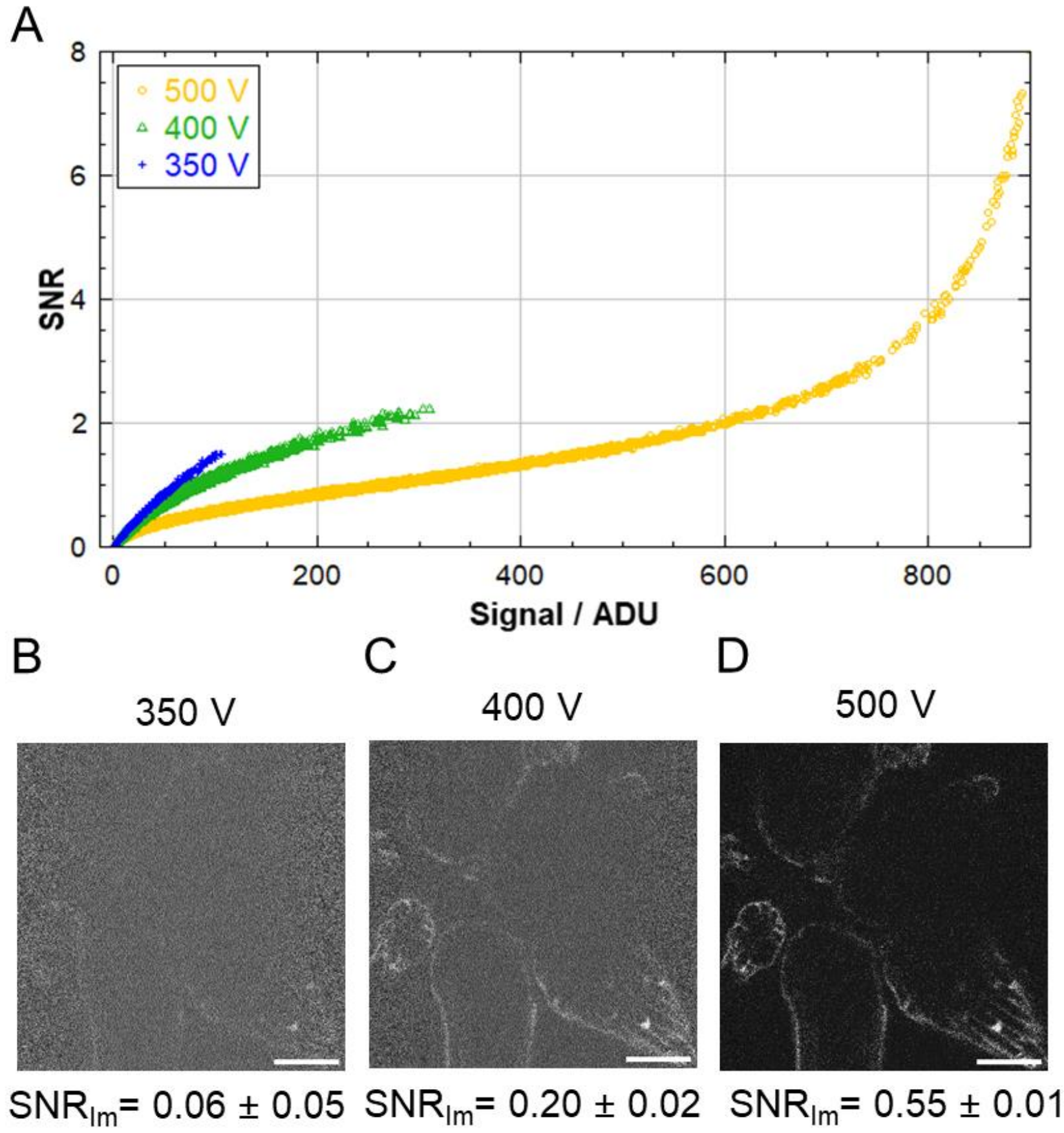

**Fig. S12:** An example of the dependence of the SNR on the PMT gain for images acquired with Olympus FVMPE-RS. (A) SNR vs. signal plots for different PMT gain and (B, C, D) examples of single frames acquired using different PMT gains (indicated in the figure) with the corresponding $SNR_{Im}$ (mean ± standard deviation from all pairs of frames in the stack); the scale bar represents 50 µm. All other acquisition settings were kept constant.